\documentstyle[12pt,aas2pp4]{article}
\def\kms{\relax \ifmmode {\,\rm km\,s}^{-1}\else \,km\,s$^{-1}$\fi}

\def\farcs{\hbox{$.\!\!^{\prime\prime}$}}

\def\secd#1.#2{ #1\farcs#2 } 

\def\td{$tD^{-1}$}

\def\mincir{\ \raise-2.truept\hbox{\rlap{\hbox{$\sim$}}\raise5.truept
    \hbox{$<$}\ }}
\def\magcir{\ \raise-2.truept\hbox{\rlap{\hbox{$\sim$}}\raise5.truept
    \hbox{$>$}\ }}
\def\gr{$^\circ$}

\def\nii{[N {\sc ii}]}

\def\heii{He {\sc ii}}
\def\oiii{[O {\sc iii}]}
\def\ha{H$\alpha$}
\def\hb{H$\beta$}

\lefthead{IC 2553 and NGC 5882}
\righthead{Corradi et al.}

\tightenlines

\begin{document}

\title{Knots in the outer shells of the planetary nebulae IC 2553 and NGC
5882\footnote{Based on observations obtained at the 3.5-m New Technology
Telescope (NTT) of the European Southern Observatory, and with the NASA/ESA
{\it Hubble Space Telescope}, obtained at the Space Telescope Science
Institute, which is operated by AURA for NASA under contract NAS5-26555.}}

\author{Romano L. M. Corradi}
\affil{Isaac Newton Group of Telescopes,\\ Apartado de Correos 321,
	 E-38780, Sta. Cruz de la Palma, Canary Islands, Spain.\\e-mail: 
	 rcorradi@ing.iac.es}

\author{Denise R. Gon\c calves, Eva Villaver, and Antonio Mampaso}

\affil{Instituto de Astrof\'{\i}sica de Canarias,\\ c. Via Lactea S/N, 
	E-38200 La Laguna, Tenerife, Spain. \\e-mail: 
	denise@ll.iac.es, villaver@ll.iac.es, amr@ll.iac.es}

\author{Mario Perinotto}

\affil{Dipartimento di Astronomia e Scienza dello Spazio, Universit\`a 
di Firenze, \\Largo E. Fermi 5, 50125 Firenze, Italy. 
\\e-mail: mariop@arcetri.astro.it}


\begin{abstract}

We present images and high-resolution spectra of the planetary nebulae
IC~2553 and NGC 5882. Spatio-kinematic modeling of the nebulae  shows that
they are composed of a markedly elongated inner shell, and of a less
aspherical outer shell expanding at a considerably higher velocity than the
inner one. 

Embedded in the outer shells of both nebulae are found several 
low-ionization knots. 
In IC~2553, the knots show a point-symmetric distribution with
respect to the central star: one possible explanation for their formation is
that they are the survivors of pre-existing point-symmetric condensations
in the AGB wind, a fact which would imply a quite peculiar mass-loss geometry
from the giant progenitor.  In the case of NGC~5882, the lack of symmetry in
the distribution of the observed low-ionization structures makes it possible
that they are the result of { in situ} instabilities.

\end{abstract}

\keywords{planetary nebulae: individual (IC 2553, NGC5882) - 
ISM: kinematics and dynamics - ISM: jets and outflows}

\section{Introduction}

Many planetary nebulae (PNe) are known to possess intriguing low-ionization
small-scale structures which usually appear in the form of low-ionization
knots, bullets, filaments, ansae, etc., embedded in the main bodies of the
nebulae or outside them. These 
structures have received a great deal of
attention in the last years (e.g., Balick et al. 1998, and references
therein) because they might provide important insights into the processes
governing PN formation and evolution, such as the role of
dynamical instabilities and clumpiness, the collimation mechanisms of the
mass outflow from single and binary AGB stars, or the occurrence of discrete
mass-loss episodes in the post-AGB phase. Presently, we are still quite far
from understanding their origin (Balick et al. 1998).

We are carrying out an observational program aimed at investigating the
properties of these low-ionization microstructures. Most of our targets were
selected by Corradi et al. (1996) by computing (\nii+\ha)/\oiii\ ratio maps
in the image catalogs of Schwarz, Corradi, \& Melnick (1992) and Gorny et
al. (1999).  In previous papers (Corradi et al. 1997, 1999, 2000a), we have
reported the results for several PNe, highlighting the occurrence of extended
highly  collimated structures (e.g., NGC 3918 and K~1-2), high-velocity
symmetrical knots (K~4-47), precessing outflows (NGC 6337 and He~2-186), and
multiple collimated ejecta (IC~4593 and Wray 17-1). From this work, it
appears that different dynamical/radiative processes are needed  to
explain the variety of low-ionization structures observed.

In this paper, we discuss the results for IC~2553 and NGC~5582, which present
an additional type of low-ionization structures:  systems
of knots located in the outer shells of the nebulae.
IC~2553 and NGC~5582 are two southern PNe which have not been extensively studied
in the past, apart from several statistical studies which are referenced
throughout the paper. A chemical study of the different morphological
structures of NGC~5882 was recently presented by Guerrero \& Manchado (1999).
In this paper we present high-resolution long-slit spectra and narrow-band
images which allow us to determine the geometry and kinematics of the main
shells of IC~2553 and NGC~5582, with particular emphasis on the
low-ionization structures contained therein.

\section{Observations}

Images and spectra of IC~2553 (PN G285.4--05.3) and NGC~5882 (PN
G327.8 +10.0) were obtained on 1996 April 28 at ESO's 3.5-m New Technology
Telescope (NTT) at La Silla (Chile), using the EMMI multimode instrument.
With the TEK 2048$^2$ CCD ESO\#36, the spatial scale of the instrument was
0$''$.27~pix$^{-1}$ both for imaging and spectroscopy.  The central
wavelength and full width at half maximum (FWHM) of the \nii\ filter used for
imaging are 658.8~nm and 3.0~nm, and those of the \oiii\ filter 500.7~nm and
5.5~nm. Further details of the observations are listed in Table~1.  As with the
spectroscopy, EMMI was used in the long-slit, high-resolution mode (Corradi,
Mampaso, \& Perinotto 1996), providing a reciprocal dispersion of
0.004~nm~pix$^{-1}$, and a spectral resolving power of
$\lambda$/$\Delta\lambda$=55000 with the adopted slit width of 1$''$.0. The
slit length was of 6~arcmin. The echelle order selected by using a broad \ha\
filter includes the \heii\ line at $\lambda$ = 656.01~nm, \ha\ at
$\lambda$ = 656.28~nm, and the \nii\ doublet at $\lambda$ = 654.81 and 658.34~nm.
The slit was positioned through the center of the nebulae at the 
position angles listed in Table~1.

We also retrieved from the {\it HST} archive images of NGC~5882 obtained on
1995 July 27 with the WFPC2 camera (PC CCD, 0$''$.0455 pix$^{-1}$) in the
F555W  filter (525.2/122.3~nm, total exposure time 24~s).
Note that the emission of NGC~5882 in this broad-band filter is expected to
be dominated by the strong \oiii\ lines at $\lambda=$ 495.9 and 500.7~nm
(cf. the spectrum in Guerrero \& Manchado 1999). However, various other
nebular lines fall within the transmission range of the filter and may be
important at specific positions in the nebula.

Images and spectra were reduced in a standard way using MIDAS and IRAF. 

\section{IC 2553}

\subsection{Morphology}

\placefigure{F-i2553i}

The \oiii\ and \nii\ NTT images of IC~2553 are presented in
Figure~\ref{F-i2553i}.  In \oiii, IC~2553 appears to be composed of
a roughly rectangular 9$''$$\times$4$''$ inner nebula  that
has two protrusions along its short axis. This inner nebula shows some
evidence for limb brightening, suggesting that it is in fact a shell.  Around
it, there is a faint outer nebula whose surface brightness rapidly
falls below the detection limit of the present observations. In \nii, the
inner shell is also clearly visible but around it there is a system of
low-ionization knots located at various position angles.  Some of them
are clearly paired and located symmetrically with respect to the central
star (labeled $AA'$ and $BB'$ in Figure~\ref{F-i2553i}).  In addition to the
knot-like features, a faint low-ionization lane extends for about 4$''$ from
the bright shell toward the north.

\subsection{Kinematics}

\placefigure{F-i2553s}

The long slit of the spectrograph was positioned through the central star at
P.A.  = $-69$\gr\ (minor axis of the inner shell), P.A. = $-1$\gr\ and
P.A. = $-142$\gr\ (along the diagonals of the cylindrical shell), intersecting
some of the low-ionization knots.  Greyscale plots of the spectra are
presented in Figure~\ref{F-i2553s}.  First, note that the knots identified in
the images are prominent in \nii\ but are often not even detected in \ha.
Second, the point-symmetry observed in the \nii\ image for some pairs of
knots is also found in the velocity data. This applies not only to the knots
$AA'$ at P.A. = $-142$\gr, but also to the bright knot $C$, which is seen
projected close to the center of the nebula, and which is found in all
spectra to have a kinematical counterpart ($C'$ in Figure~\ref{F-i2553s}) with
opposite symmetrical radial velocities.

Probably caused by the large thermal width of \ha, the corresponding
position--velocity plots do not allow us to separate kinematically the
different morphological components of the nebula. This is instead clearly
achieved in the \nii\ spectra, where one can immediately see that the
emission from the inner bright shell is clearly separated from that of the
(higher velocity) outer nebula and knots.  To get an insight into the
geometrical, orientational and kinematic properties, we have applied the same
kind of spatio-kinematic analysis as for NGC~3918 in a previous paper 
(Corradi et al. 1999).  First, we measured by multiple-Gaussian
fitting the \nii\ radial velocities in the different regions of the nebula at
all slit position angles. That was done with 2-pixel spatial binning along
the slit for the inner shell and for the regions which are spatially
extended, while we used larger binnings ($1''$ to $3''$) to measure the
average velocity of each knot. The \nii\ velocities are shown in
Figure~\ref{F-i2553fit}: small dots refer to the smaller binnings, large dots
to the knots' velocities. Note that, along the short axis of the nebula
(P.A. = $-69$\gr), the \nii\ spectrum clearly reveals the existence of two
separate shells, the outer one expanding at a faster velocity than the inner
one.

We first analyzed the velocity field of the inner shell.  Its
kinematic, geometrical and orientational parameters were derived using the
heuristic spatio-kinematic model of Solf \& Ulrich (1985). We fitted the \nii\
position--velocity plots and the shape of the inner shell with an
axisymmetric model in which the nebular expansion velocity, $V_{\rm exp}$,
increases from the equatorial plane toward the polar axis, following the
equation (Solf \& Ulrich 1985):
\begin{equation} 
V_{\rm exp}(\phi)=V_{\rm e}+(V_{\rm p}-V_{\rm
e})\sin^\gamma(|\phi|), 
\end{equation} 
where $\phi$ is the latitude angle (0$^\circ$ in the equatorial plane,
90$^\circ$ in the polar directions), $V_{\rm p}$ and $V_{\rm e}$ are the
polar and equatorial velocities, and $\gamma$ is a shape parameter.  The other
quantities involved in the model are the heliocentric systemic velocity,
$V_\odot$, the inclination, $i$, of the nebula (the angle between the polar axis
and the line of sight), and the product \td\ containing the inseparable
effects on the apparent nebular size due to the distance, $D$, and to the
kinematic age, $t$.

To avoid the knots and better highlight the shape of the inner shell, we have
used for the fit the \oiii\ image (the contour plot in the bottom-right panel of
Figure~\ref{F-i2553fit}). The best fit of both the shape and kinematics of the
inner shell (the thick continuous line in Figure~\ref{F-i2553fit}) is obtained with
the following parameters: $i$ $=$ 78\gr, \td\ $=$ 0.55~yr~pc$^{-1}$, $V_{\rm
p}$ $=$ 35~\kms, $V_{\rm e}$ $=$ 17~\kms, $\gamma$ $=$ 2, and V$_\odot$  $=$ 30~\kms.
The fit of the spectra at P.A. = $-142$\gr\ and P.A. = $-69$\gr\ is good, as 
is that of the \oiii\ shape if the two equatorial protrusions are not
considered (in the \nii\ image of Figure~\ref{F-i2553i}, they in fact appear to
be equatorial knots separated from the inner shell).  Some deviations from
the model are visible for the spectrum at P.A. = $-1$\gr\ along the polar
directions, in connection with the northern extended lane and on the
symmetrically opposite side.  They probably reflect a peculiar kinematics in
these regions.

\placefigure{F-i2553fit}

In spite of these local deviations, the overall fit is fairly good, and 
we conclude that the inner nebula of IC~2553 is an elongated shell
seen slightly inclined to the plane of the sky and with a slight equatorial
waist, the polar velocities being about twice the equatorial ones.

As noted above, the short-axis spectrum clearly highlights the existence
of an outer, higher-velocity shell. That spectrum indicates in fact the
existence of an equatorial torus detached from the inner shell and expanding
with a velocity twice as large (35~\kms, assuming the same polar axis as
for the inner shell). Since the low-ionization knots also have a higher
velocity than the inner shell, we investigated the possibility that the
equatorial torus and the knots are in fact part of an outer shell expanding
with larger velocities than the inner one.  We  then repeated the
spatio-kinematic modeling, looking for a solution for the outer shell that
is geometrically not too different from that of the inner one. A reasonable
fit to the short axis and knot velocities, and to the \oiii\ surface
brightness of the outer shell is indicated by the dashed lines in
Figure~\ref{F-i2553fit}, and corresponds to the following parameters:
$i$ $=$ 78\gr\ (fixed from the solution for the inner shell),
\td\  $ =$ 0.50~yr~pc$^{-1}$, $V_{\rm p}$ $=$ 58~\kms, $V_{\rm e}$ $=$ 35~\kms,
$\gamma$ $=$ 4, and V$_\odot$ $=$ 33~\kms.  This means that the outer 
shell would
be slightly more spherical than the inner one (as also suggested by the image
plot contours), and would have equatorial velocities twice as large. Also,
and more importantly, the whole system of low-ionization knots would be contained
within this outer shell and would not be moving with peculiar velocities
within it.

From the average of the values derived from the two fits, we adopt a
heliocentric systemic velocity for IC~2553 of $31$$\pm$$4$~\kms\, in good
agreement with the value $37$$\pm$$6$~\kms, which is the weighted mean of
the values in the compilation by Durand, Acker, \& Zijlstra (1998). 

Finally, the low-ionization lane extending toward the north has low 
radial velocities (see the upper-left panel of Figure~\ref{F-i2553fit}). If we
assume that it is directed roughly along the polar axis of the shells, then
its deprojected expansion velocity would be increasing from inside to outside
from $\sim$ 30 to $\sim$ 50~\kms, which is, as in the case of the knots, similar to
the general expansion velocity of the outer shell gas in these polar regions.

\section{NGC5882}

\subsection{Morphology}

\placefigure{F-n5882i}

The {\it HST} and the NTT images of NGC 5882 are presented in
Figure~\ref{F-n5882i}.  In the {\it HST} image, NGC~5882 appears to be
composed of a bright elliptical inner shell measuring
11$''$$\times$6$''$ surrounded by a fainter more spherical outer shell
with a diameter of about 15$''$.  The surface of the inner rim seems
to be composed of several bubble-like structures.  
Note that the central star is not at the center of symmetry of the rim, but
is displaced toward its western side. 
We do not discuss further this
deviation from axisymmetry, which may be the result of large scale
instabilities in the mass loss process and/or of the interaction with a
binary companion (cf. Soker 1999), and might therefore provide additional
information about the formation mechanism of the markedly elliptical rim.

The outer shell might in
turn be subdivided into two regions with different surface brightnesses,
the inner, brighter one having a sharp outer edge and a more elongated
shape than the outer one. The present data do not allow us to
discern whether they are indeed two distinct shells, and
therefore in  future we shall refer to the whole region as the
``outer shell''.

In the low-ionization emission of \nii, several knots appear in the outer
shell. The brightest ones, located in the north-west quadrant (indicated
by arrows in Figure~\ref{F-n5882i}), are also visible, albeit much fainter, in
the {\it HST} images, in which they
 are resolved as complex systems of filaments and
knots lying at the sharp edge of the outer shell. Similar low-ionization
structures are seen in the PN NGC 7662 (Balick et al. 1998, who called them
``tail-head microstructures'': see their Figures 1, 2 and 5).  At
variance with IC~2553, the low-ionization structures of NGC~5882 do not show
any clear symmetry around the center.

The faint extended halo with a radius of up to 160$''$ that is known to
surround NGC~5882 (Guerrero \& Manchado 1999) is only weakly revealed by our
NTT images and is not further discussed in this paper.

\subsection{Kinematics}

\placefigure{F-n5882s}

The long slit of the spectrograph was positioned through the central
star at P.A. = $-85$\gr\ (approximately the minor axis of the inner
shell), P.A. = $-19$\gr,  and 
P.A. = $-45$\gr, intersecting the low-ionization knots found in the
\nii\ image.  Greyscale plots of the spectra are presented in
Figure~\ref{F-n5882s}.  There are several similarities with
IC~2553. First, in both \ha\ and \nii\ we observe an elliptical
kinematic figure typical of an expanding, hollow ellipsoidal shell
identified with the inner shell of NGC~5882. Second, the knots in the
outer shell are clearly visible only in  \nii, and 
generally have larger radial velocities than the inner shell.

We then attempted the kind of spatio-kinematic modeling used for IC~2553.
Radial velocities were measured by means of multiple-Gaussian fitting of the
spectra at the different slit position angles.  For NGC~5882, beside the
\nii\ velocities (filled circles in Figure~\ref{F-n5882fit}), we also display
the \ha\ velocities (empty circles) which present a slightly more regular
velocity pattern than \nii.  The fit in Figure~\ref{F-n5882fit} is not as good
as in the case of IC~2553 but is still fair enough to draw general conclusions
about the geometrical, orientation and kinematical properties of the inner
shell of NGC~5882 and  its low-ionization knots. First, the inner shell is
found to be an ellipsoid only slightly inclined to the plane of the sky.  The
parameters obtained by fitting the \ha\ velocities and the shape in the {\it HST}
image are the following: $i$ $=$ 80\gr ($\pm$10\gr), \td $=$ 0.65~yr~pc$^{-1}$,
$V_{\rm p}$ $=$ 38~\kms, $V_{\rm e}$  $=$22~\kms, $\gamma$  $=$2.8, and
V$_\odot$ $=$ 12~\kms. The \nii\ velocity field of the inner shell is more
irregular, but on  average the expansion velocity in this ion is only
slightly larger than for \ha.

\placefigure{F-n5882fit}

Previous kinematic studies of NGC~5882 quoted an expansion velocity of
12.5~\kms\ in \hb\ and \oiii\ (Banerjee et al. 1990), 16.5~\kms\ in \ha, and
23.5 in \nii\ (Ortolani \& Sabbadin 1985).  Note, however, that those studies
did not spatially resolve the nebula, while we have shown that the expansion 
velocity is not isotropic, but varies with direction.

The heliocentric systemic velocity of NGC~5882 derived from our modeling is
 $12$$\pm$$4$~\kms, which is comparable with  values found by previous
authors: 9.8~\kms\ (Bianchi 1992), and 7.7~\kms\ (Campbell et al. 1981). 
However, Acker (1978) quote 21~\kms\ for this object. 

As far as the outer shell of NGC~5882 is concerned, the \nii\ radial
velocities of the knots and of other regions in this shell suggest
that also in this object expansion velocities in the outer shell are
systematically larger than in the inner one. The morphology and
velocity field are quite irregular and fragmentary, so that we did not
attempt to make a spatio-kinematic fitting as for the inner
rim. Nevertheless, the present data suggest that we might well be in
the same situation as for IC~2553, the knots being part of the outer
shell which would be expanding with a velocity between
10 and 20~\kms\ larger than that of the inner shell.

Note that some of the knots of NGC~5882 present a complex kinematic
pattern. The structure at P.A. = $-85$\gr\ at a distance from 2.4 to 4.5
arcsec from the center (indicated by a circle in Figures~\ref{F-n5882i} and
\ref{F-n5882s}) is in fact split, in the position--velocity plot, into four
subcomponents, each pair having a velocity difference of about 15~\kms. In
the {\it HST} image, however, this structure corresponds to a diffuse region of
faint emission within the outer shell and is not associated with any obvious
morphological feature.

\section{Discussion}

According to the present data, IC~2553 and NGC~5882 share several properties.
For the following discussion, we assume that both nebulae are indeed composed
of an elongated inner shell, as well as of a higher-velocity, more spherical outer
shell. The latter, when observed in a low-ionization emission line, contains
several knots which do not show evidence of moving with peculiar
velocities compared to the general expansion motion of the outer shell.

The main differences between the two PNe are the following. First, the inner
and outer shells of IC~2553 are more aspherical than those of
NGC~5882. Second, the velocity field of IC~2553 is more regular, and thus our
modeling gives clearer evidence that the outer shell is expanding
faster than the inner one, and that the knots participate in the
expansion velocity of that shell. Third --- and probably the most important
difference --- the system of knots of NGC~5882 does not show the
point-symmetry displayed by the knots of IC~2553.

A two-shell configuration is common in PNe.  A large fraction of PNe (see,
for example, Chu et al. 1987; Stanghellini \& Pasquali 1995) have in fact outer shells
which are ``attached to'' or ``detached from'' their bright inner rims.
Theoretically, a double-shell configuration is expected to develop as a
consequence of the combined action of photo-ionization and fast vs. slow wind
interaction; Marten \& Sch\"onberner (1991) and Mellema (1994) showed that
the outer shell is formed by the ionization front while the inner shell is
swept-up by the fast wind. Because of the action of the ionization front, the
outer shell can acquire a velocity larger than that of the inner
shell. This is indeed observed in several PNe (cf. Chu 1989; Mellema 1994;
Guerrero, Villaver, \& Manchado 1998).  According to Mellema (1994), a common
characteristic of the PNe with outer shells faster than the inner ones is
that they are relatively young.  Later in the evolution, in fact, the inner
shell accelerates and ends up moving with the same or an even higher velocity
than the outer shell (Mellema 1994; see also Corradi et al. 2000b).  

The distance of IC~2553 is uncertain, and determinations in the literature
span from 0.5 to 4~kpc (Zhang \& Kwok 1993; Mal'kov 1997). Even adopting the
larger distances, the kinematic age of IC~2553 as derived from the above
spatio-kinematic modeling is smaller than some 2000~yr. Although
there could be some underestimate of the real dynamical age (Sch\"onberner et
al. 1997), this confirms that IC~2553 is not an evolved PN. The temperature
of its central star is slightly less than 100 000~K (Stanghellini, Corradi,
\& Schwarz 1993; Mal'kov 1997), and both the 1-D simulations in Mellema
(1994) and in Corradi et al. (2000b) show that at this relatively evolved
stage of post-AGB evolution the velocity of the outer shell can still be
greater than that of the inner one.  The same reasoning also holds for
NGC~5882.  The distances estimated for these objects span from 1.1 to 
2.4~kpc (Cahn
1976; Sabbadin 1984; Ortolani \& Sabbadin 1985), corresponding to a
kinematic age for the inner shell smaller than 1500~yr,  thus indicating
that this PN is also relatively young. The temperature of its central star is
 $\sim$ 70000~K (Zhang \& Kwok 1991, 1993).  Thus the observed properties
of IC~2553 and NGC~5882 are consistent with the idea that they are PNe found
in an evolutionary phase in which the outer shell is expanding faster than
the inner one.  Regarding the fact that the inner shells of both nebulae are
more aspherical than the outer ones, this might be a direct effect of the
wind--wind interaction, in particular of the anisotropical expansion of the
associated shock front.  Note that all the gas-dynamical models referred above
were developed for spherical geometry, while IC~2553 and NGC~5882 are
markedly asymmetrical; thus we are assuming that the same behavior as
predicted/observed for spherical PNe also qualitatively applies to (at least
some) asymmetrical objects.

Guerrero et al. (1998), however, found that the ratio between the expansion
velocities of outer and inner shells would increase with the ellipticity of
the inner shell, rather than being  purely an effect of age. 
NGC 5882 and 
IC~2553 particularly fit well into this idea; when added to the sample of Guerrero et
al. (1998), they would possess among the most aspherical inner shells and the
highest ratios between the expansion velocities of the outer and inner
shells. This is a very interesting issue to be investigated further in
future, especially from a theoretical point of view.

What is the origin of the low-ionization knots found in the outer shells of
IC~2553 and NGC~5882? The present data suggest that the expansion velocity of
the knots is not peculiar with respect to that of their environments (the
outer shell gas). Thus, according to the definition by Balick et al. (1993),
the knots are not genuine FLIERs, and are not kinematically younger than the
surrounding gas. This means that they are not material recently ejected by
the central star (unless they have been dramatically slowed down).  In
addition, since they are found in a region external to the shock driven by
the fast post-AGB wind into the slow AGB wind, we conclude that their
formation is not related to the interacting-winds processes which drive the
evolution of the bright rim of the nebula.

Instead, it is possible that the formation of the knots is due to ionization
effects, which are the main cause of  the dynamical evolution of the
outer shell. This idea would indeed be consistent with the finding that the
knots do not have peculiar velocities. However, at least in the case of
IC~2553, one has also to account for the fact that various knots are found in
symmetrical pairs with respect to the central star. This symmetry rules out
the possibility that the knots are formed by in situ instabilities (for
instance, thermal instabilities being enlarged by the expansion of the
ionization front), since instabilities would not give rise to any clearly
symmetrical pattern.  The possibilities that remain are the following: i)
the knots are pre-existing (fossil) point-symmetric structures in the AGB
wind that have survived (or have even been exacerbated) until the present age 
to dynamical and
ionization effects (Garc\'\i a-Segura \& Franco 1996; Mellema et al. 1998;
Soker 1998); ii) they are the traces, on the surface of the outer shell, of
the passage of high-velocity material ejected from the central star in a
point-symmetric fashion.

The idea of fossil knots, e.g., condensations in the AGB wind, may find
support in the recent observations of some AGB stars, such as TT~Cyg for
which Olofsson et al. (2000) suggest a clumpy structure with a typical size
of \mincir10 $^{16}$~cm. In the PN phase, these AGB clumps will interact with
the ionization front. Along these lines, several models were previously
developed to explain the formation of low-ionization microstructures in PNe
(e.g., Mellema et al. 1998; Soker 1998; Soker \& Reveg 1998). The main
physical process to be taken into account here is the disturbance due to
the ionization of a pre-existing clump.  The timescale for the ionization of the
entire clump depends on several parameters, such as  size, density and
temperature, the ionizing flux, etc. The analytical and numerical
calculations by Mellema et al. (1998) consider a neutral clump with a size of
2$\cdot$10$^{16}$~cm being photo-evaporated by the ionizing radiation from
the central star; the corresponding timescale for ionization would be of the
order of $10^3$~yr.  While being ionized, the cross-section of such a clump
would increase with an expansion time of the order of $300$~yr (Soker \& Reveg
1998). Thus, after  ionization, the volume of the clump would increase, and
its density decrease, to end up as a region with a density not strongly
different from that of the surrounding gas.

In the case of IC~2553, the fact that the knots are barely resolved, coupled
with the large uncertainty in the distance to the object, allows us to
put only an upper limit of 6$\cdot$10$^{16}$~cm to the linear size of the largest
observed low-ionization knots. Even if they were smaller, of the sizes
considered by Mellema et al. (1998) or derived by Olofsson et al. (2000), the
 above modeling clearly shows that they would survive  photo-ionization
effects for at least $10^3$~yr, which is consistent with the kinematic post-AGB
age of IC~2553 computed from our observations.

If the low-ionization knots in this nebula are the survivors of pre-existing
structures in the original Mira wind, this implies a peculiar AGB mass-loss
geometry in which mass is ejected in very different directions in a
point-symmetric fashion.  Such point-symmetric mass ejection might be
related to non-radial pulsation of the red giant, and/or to the interactions
in a binary system (Soker \& Harpaz 1992; Livio \& Pringle 1997). It should
also be calculated at some stage whether these AGB condensations can be
accelerated by the ionization front to the considerable velocities that we
have measured.

The second possibility for the formation of these pairs of knots is that they
are the imprints of the interaction on the outer shell of some high-velocity
material recently ejected by the central star (but not the ejecta themselves,
given the relatively low velocity observed). If so, one has still to explain
the special geometry of the ejecta required to produce the observed
point symmetry. Should the ejecta be a system of high-velocity clumps, 
neither is
it clear whether they would not leave certain signatures while crossing
the inner rim, or even be evaporated and disrupted during this interaction,
nor why we do not now see them at further distances from the central
star. For these reasons, and implying a progenitor mass-loss history even
more peculiar than in the case they were the survivors of pre-existing
structures in the AGB wind, we consider the latter hypothesis to be
 more plausible.

As far as NGC 5882 is concerned, since the constraint of a point-symmetric
knot distribution does not apply here,  the simpler hypothesis
that the low-ionization structures in the outer shell are  in situ
instabilities is also indeed viable.  These instabilities would be related 
to the
expansion of the slow wind through the ISM (or through circumstellar matter
from previous mass-loss episodes), or to the expansion of the ionization
front.  In principle, with the appropriate velocity regimes the expanding
slow wind in a non-homogeneous medium could be the source of Rayleigh--Taylor,
Kelvin--Helmholtz and Vishniac instabilities (Garc\'\i a-Segura \& Franco
1996).

\section{Summary and conclusions}

The origin of the low-ionization microstructures which are observed in a
large number of PNe is intriguing. From the data
collected in this paper and in previous ones (see \S\ 1), it turns out that
they appear with so wide a variety of morphological, kinematic, physical
and topographical properties, that one can hardly think of a common physical
mechanism able to account for all of them.
 
In this article, we have focused our attention on the nature of low-ionization
knots in the outer shells of PNe (see also Balick et al. 1998). Being found
within nebular regions which, supposedly, have not yet been 
affected by the action
of the fast AGB wind which shapes the bright rims of PNe, some of the
mechanisms proposed to explain their origin can be excluded (such as the
occurrence of instabilities in the fast vs. slow winds interaction). One attractive option  is that they are the product of
instabilities in the interaction of the AGB wind with the ionization front,
the ISM, or previous AGB ejecta.  This might well apply in the case of
NGC~5882, but the point symmetry shown by the knots in IC~2553 indicates instead
 that they are the remnants of pre-existing, symmetric
condensations in the AGB wind.  Clumpy AGB winds have been observed in
molecular lines (Olofsson et al. 2000), but the additional requirement of a
point-symmetric distribution of the clumps would imply a very specific 
mass-loss geometry from the progenitor star. Understanding the physical processes
causing this kind of mass loss in AGB stars clearly deserves further
observational and theoretical study.

\section{Acknowledgments}

The work of RLMC, EV, and AM is supported by a grant of the Spanish DGES
PB97-1435-C02-01, and that of DRG by a grant from the Brazilian Agency
Funda\c c\~ao de Amparo \`a Pesquisa do Estado de S\~ao Paulo (FAPESP;
98/07502-0).

\clearpage

\figcaption[ ]{The NTT images of IC 2553, on a logarithmic intensity scale. 
The  positions of the slit used for spectroscopy are indicated by short 
lines on either side of the nebula. Point-symmetric low-ionization knots
are labeled with capital letters (see text).
\label{F-i2553i}}

\figcaption[ ]{
The NTT long-slit spectra of IC~2553 on a logarithmic scale. 
\label{F-i2553s}}

\figcaption[ ]{Kinematic modeling for IC~2553 using the description in
Solf \& Ulrich (1985). The left and upper-right panels show the \nii\
heliocentric velocities (dots) at the three slit positions through IC~2553:
large dots indicate the velocity of the low-ionization knots. In the
bottom-right box, the shape of the nebula is shown as plot contours of the
\oiii\ image, rotated in order to have the short (equatorial) axis of the
nebula aligned along the horizontal direction. Successive levels are 
incremented by a factor of $\protect \sqrt 2$. Model fits for the inner and outer shells are the
solid and dashed lines, respectively.
\label{F-i2553fit}}
\figcaption[ ]{The {\it HST} and NTT images of NGC 5882 on a linear scale. 
Slit positions are indicated by short lines, while arrows mark the position
of several low-ionization knots. The circle to the west indicates a region
of peculiar kinematics (see text).
\label{F-n5882i}}

\figcaption[ ]{
The NTT long-slit spectra of NGC~5882 on a logarithmic intensity scale. 
\label{F-n5882s}}

\figcaption[ ]{
Kinematic modeling for NGC 5882. The representation is
as in Figure~\ref{F-i2553fit}, except that beside 
the \nii\ radial velocities (full circles) we also display the 
\ha\ ones (empty circles). Large dots indicate the \nii\ velocities of the
low-ionization knots. In the bottom-right box, the shape of the nebula
is shown as plot contours of the {\it HST} image; successive levels are
incremented by a factor of $\protect \sqrt 2$. Model fits for the inner
shell are plotted as solid lines.
\label{F-n5882fit}}

\newpage
\begin{deluxetable}{lllc}
\tablenum{1}
\tablewidth{30pc}
\tablecaption{Log of observations}
\tablehead{
\multicolumn{4}{c}{\bf\it Images} \\
\multicolumn{1}{l}{Object}& 
\multicolumn{1}{l}{Telescope} & 
\multicolumn{1}{l}{Filter (exposure time, sec)} &
\multicolumn{1}{c}{Seeing}}
\startdata
IC 2553  & NTT & \nii\ (180), \oiii\ (120)    & 0$''$.9  \nl
NGC 5882 & NTT & \nii\ (600)                  & 0$''$.8  \nl
         & HST &  F555W (24)                  &      \nl
         &     &                              &      \nl
\multicolumn{4}{c}{\bf\it Long--slit spectra} \nl
       &           & P.A. (exposure time, sec) &        \nl
\hline \nl
IC 2553  & NTT & $-142$\gr\ (600), $-69$\gr\ (600), $-1$\gr\ (600)& 0$''$.9 \nl
NGC 5882 & NTT & $-85$\gr\ (600), $-45$\gr\ (600), $-19$\gr\ (600) & 0$''$.8 \nl
\enddata
\end{deluxetable}

\end{document}